\documentclass[aps,prl,twocolumn,superscriptaddress]{revtex4-1}
\usepackage{graphicx}
\usepackage{dcolumn}
\usepackage{bm}

\begin{document}


\title{First observation of single beta decay of $^{96}$Zr}


\author{A.S.~Barabash} 
\affiliation{National Research Centre "Kurchatov Institute", Moscow, 123182, Russia}

\author{S.~Evseev}
\affiliation{Laboratory of Nuclear Problems, Joint Institute for Nuclear Research, 6 Joliot-Curie, Dubna, 141980, Russia}

\author{D.~Filosofov}
\affiliation{Laboratory of Nuclear Problems, Joint Institute for Nuclear Research, 6 Joliot-Curie, Dubna, 141980, Russia}

\author{Yu.M.~Gavrilyuk}
\affiliation{Institute for Nuclear Research of the Russian Academy of Sciences, 7a Prospect 60-letiya Oktyabrya, Moscow, 117312, Russia}

\author{A.M.~Gangapshev}
\affiliation{Institute for Nuclear Research of the Russian Academy of Sciences, 7a Prospect 60-letiya Oktyabrya, Moscow, 117312, Russia}

\author{N.~Gorshkov}
\affiliation{Laboratory of Nuclear Problems, Joint Institute for Nuclear Research, 6 Joliot-Curie, Dubna, 141980, Russia}

\author{V.V.~Kazalov}
\affiliation{Institute for Nuclear Research of the Russian Academy of Sciences, 7a Prospect 60-letiya Oktyabrya, Moscow, 117312, Russia}
\affiliation{Berbekov Kabardino-Balkarian State University, Nalchik, Russia}
\affiliation{Laboratory of Nuclear Problems, Joint Institute for Nuclear Research, 6 Joliot-Curie, Dubna, 141980, Russia}

\author{S.~Kazartsev}
\affiliation{Laboratory of Nuclear Problems, Joint Institute for Nuclear Research, 6 Joliot-Curie, Dubna, 141980, Russia}
\affiliation{Institute for Nuclear Research of the Russian Academy of Sciences, 7a Prospect 60-letiya Oktyabrya, Moscow, 117312, Russia}

\author{T.~Khussainov}
\email{khusainov@jinr.ru}
\affiliation{Laboratory of Nuclear Problems, Joint Institute for Nuclear Research, 6 Joliot-Curie, Dubna, 141980, Russia}
\affiliation{Institute of Nuclear Physics of the Ministry of Energy of the Republic of Kazakhstan, 1 Ibragimov Street, Almaty, 050032, Kazakhstan}

\author{V.V.~Kuzminov}
\affiliation{Institute for Nuclear Research of the Russian Academy of Sciences, 7a Prospect 60-letiya Oktyabrya, Moscow, 117312, Russia}

\author{A.~Lubashevskiy}
\affiliation{Laboratory of Nuclear Problems, Joint Institute for Nuclear Research, 6 Joliot-Curie, Dubna, 141980, Russia}
\affiliation{Institute for Nuclear Research of the Russian Academy of Sciences, 7a Prospect 60-letiya Oktyabrya, Moscow, 117312, Russia}
\affiliation{Lebedev Physical Institute of the Russian Academy of Sciences, 53 Leninskiy Prospect, Moscow, 119991, Russia}

\author{D.V.~Ponomarev}
\affiliation{Laboratory of Nuclear Problems, Joint Institute for Nuclear Research, 6 Joliot-Curie, Dubna, 141980, Russia}
\affiliation{Institute for Nuclear Research of the Russian Academy of Sciences, 7a Prospect 60-letiya Oktyabrya, Moscow, 117312, Russia}
\affiliation{Lebedev Physical Institute of the Russian Academy of Sciences, 53 Leninskiy Prospect, Moscow, 119991, Russia}

\author{S.~Rozov}
\affiliation{Laboratory of Nuclear Problems, Joint Institute for Nuclear Research, 6 Joliot-Curie, Dubna, 141980, Russia}

\author{N.~Temerbulatova}
\affiliation{Laboratory of Nuclear Problems, Joint Institute for Nuclear Research, 6 Joliot-Curie, Dubna, 141980, Russia}
\affiliation{Institute of Nuclear Physics of the Ministry of Energy of the Republic of Kazakhstan, 1 Ibragimov Street, Almaty, 050032, Kazakhstan}

\author{S.~Vasilyev}
\affiliation{Laboratory of Nuclear Problems, Joint Institute for Nuclear Research, 6 Joliot-Curie, Dubna, 141980, Russia}

\author{E.A.~Yakushev}
\affiliation{Laboratory of Nuclear Problems, Joint Institute for Nuclear Research, 6 Joliot-Curie, Dubna, 141980, Russia}

\author{V.I.~Yumatov}
\affiliation{National Research Centre "Kurchatov Institute", Moscow, 123182, Russia}




\date{\today}

\begin{abstract}
The single beta decay of $^{96}$Zr has been detected for the first time using a 211 cm$^3$ low-background HPGe detector and an external source consisting of two samples of enriched zirconium (atomic fraction of $^{96}$Zr is 88.28\%, total mass is 140.65 g). During the search for the $\beta$ decay of $^{96}$Zr, the $\beta$ decay of the daughter nucleus $^{96}$Nb to the excited states of $^{96}$Mo has been observed. The $\gamma$-ray cascade produced by the $^{96}$Mo nucleus while de-exciting to the ground state has been detected with the HPGe detector. The experiment has been carried out at the Baksan Neutrino Observatory. It has produced 12625.34 h of data. The half-life of the single beta decay of $^{96}$Zr is measured to be $T_{1/2} = [2.27^{+0.53}_{-0.36}(stat) \pm 0.27(syst)]\times10^{20}$ yr. 

\end{abstract}

\keywords{beta decay, $^{96}$Zr}

\maketitle

The search for neutrinoless double beta decay ($0\nu\beta\beta$) is one of the most urgent problems in nuclear and particle physics (see recent reviews \cite{AGO23,GOM23,ADA22,SIM21}). The discovery of this process will automatically lead to two fundamental conclusions: 1) the lepton number $L$ is violated, and 2) the neutrino is a Majorana particle. In addition, it will provide information on such fundamental issues as the absolute neutrino mass scale, the type of hierarchy and the CP violation in the lepton sector. The registration of the process with Majoron emission will lead to the discovery of a new elementary particle, Majoron, one of the candidates for dark matter \cite{BER93,ROT93,GAR17,BRU19,MAN23}. All this will have major consequences in physics and astrophysics. The study of different types of isotopes and decays will provide information about the mechanism of $0\nu\beta\beta$ decay.

One of the main problems is the low accuracy in calculating the values of nuclear matrix elements (NME) for $0\nu\beta\beta$ decay. It is generally accepted that the accuracy of such calculations is approximately a factor of 3. This leads to a large spread of values of the Majorana neutrino effective mass determined on the basis of experimental data. The attempts to improve the accuracy of NME calculations have been ongoing for a long time. The main hopes here lie, on the one hand, in improving the theory and, on the other hand, in using the existing experimental data on two-neutrino double beta decay ($2\nu\beta\beta$), which proceeds to both the ground and excited states of the daughter nucleus. The comparison of experimental data with calculations will help us to create an adequate theory for both $2\nu\beta\beta$ and $0\nu\beta\beta$ decays.

One of the most promising nuclei for studying double beta decay is the $^{96}$Zr nucleus since it has one of the highest 2$\beta$ transition energies both for the transition to the ground state ($Q_{\beta\beta}$ = 3356.1 keV) and to the first 0$^+_1$ excited state ($Q_{\beta\beta}$ = 2208.0 keV) of the daughter nucleus. This nucleus is quite unusual: it can decay both via the $2\nu\beta\beta$ decay channel and via the $\beta$-decay channel ($Q_{\beta}$ = 164 keV). The ordinary $\beta$ decay is highly suppressed due to the large difference in quantum numbers of the initial and final states and the relatively low transition energy. 

\begin{figure*}[htb]
\begin{center}
\includegraphics[width=0.7\textwidth]{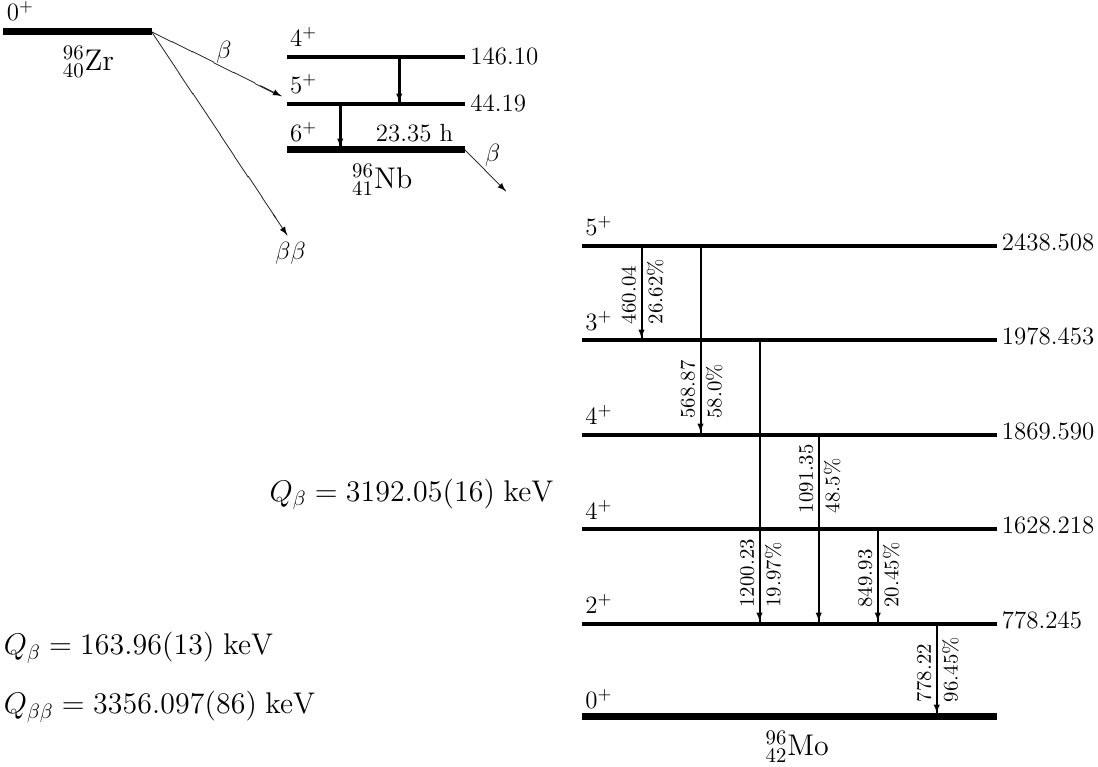}
\caption{Simplified scheme of $\beta$ decays $^{96}$Zr $\to$ $^{96}$Nb $\to$ $^{96}$Mo. The most intense $\gamma$ lines with energies and shares in decay (in per cent) are shown. The energies of the levels and emitted $\gamma$ quanta are taken from \cite{NDS08} (data on the $\beta$ decay of $^{96}$Nb). The Q-values are from \cite{ALA16}.}
\label{fig:1}
\end{center}
\end{figure*} 

Among all nuclei, the candidates for $2\beta$ decay, only $^{48}$Ca can also decay via the single $\beta$-decay channel. The $2\nu\beta\beta$ decay of $^{96}$Zr to the ground state of the daughter nucleus has been detected, and its half-life is well known, $T_{1/2} = (2.3 \pm 0.2)\times10^{19}$ yr \cite{BAR20}. However, the ordinary $\beta$-decay of $^{96}$Zr has not yet been registered, and currently the best experimental limit is $3.8\times10^{19}$ yr \cite{ARP94}. If this decay is also registered, there will be a unique context for the theoretical study of this nucleus. The calculations must be consistent for two decays, the 4-fold forbidden ordinary $\beta$ decay and the transition to the ground state of the daughter nucleus via $2\nu\beta\beta$ decay. This allows us to hope for more accuracy in determining the parameters of the theory, which will improve theoretical schemes and the precision of NME calculations for both $2\nu\beta\beta$ decay and $0\nu\beta\beta$ decay. The study of the $^{96}$Zr system will also clarify the situation with the ''quenching'' of the axial vector constant $g_A$ \cite{ENG17,EJI19,SUH19}. At present, it is not clear whether this is a real decrease in the effective value of $g_A$ in the nuclear substance, or whether it is related again to the problem of calculating the NME \cite{GYS19}. If the $2\nu\beta\beta$ decay of $^{96}$Zr into the 0$^+_1$ excited state of $^{96}$Mo is also registered (the best modern limit is $3.1\times 10^{20}$ yr \cite{FIN15}), the situation will become even more intriguing: the theory must provide a consistent description of three different decays of $^{96}$Zr. Currently, this type of decay has been registered for two isotopes, $^{100}$Mo and $^{150}$Nd (see \cite{BAR20}). $^{96}$Zr is the next most promising candidate for registering this decay \cite{BAR90}.

\begin{figure*}[htb]
\begin{center}
\includegraphics[width=0.9\textwidth]{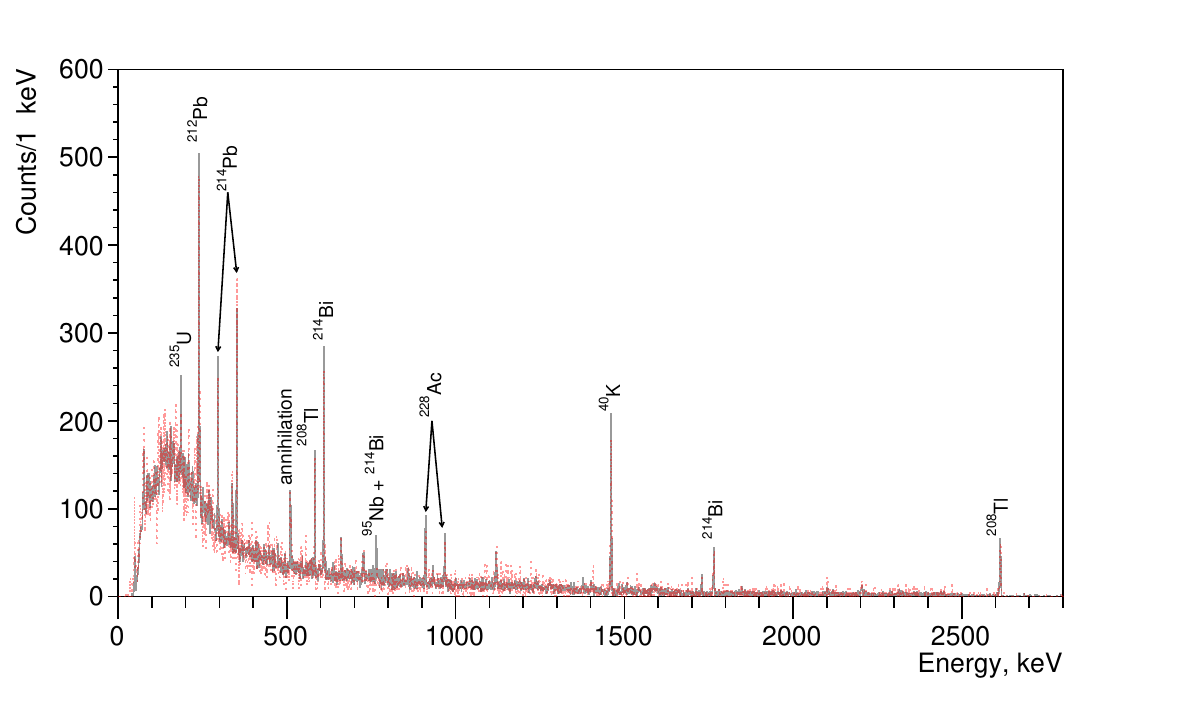}
\caption{Spectra of $\gamma$ quanta in the energy region of 0-2800 keV. The black line is the measurement with enriched Zr samples during 12625.34 h. The red (dotted) line is the background measurement during 2896.98 h, normalized to 12625.34 h. }
\label{fig:1}
\end{center}
\end{figure*} 

\begin{figure*}[htb]
\begin{center}
\includegraphics[width=0.9\textwidth]{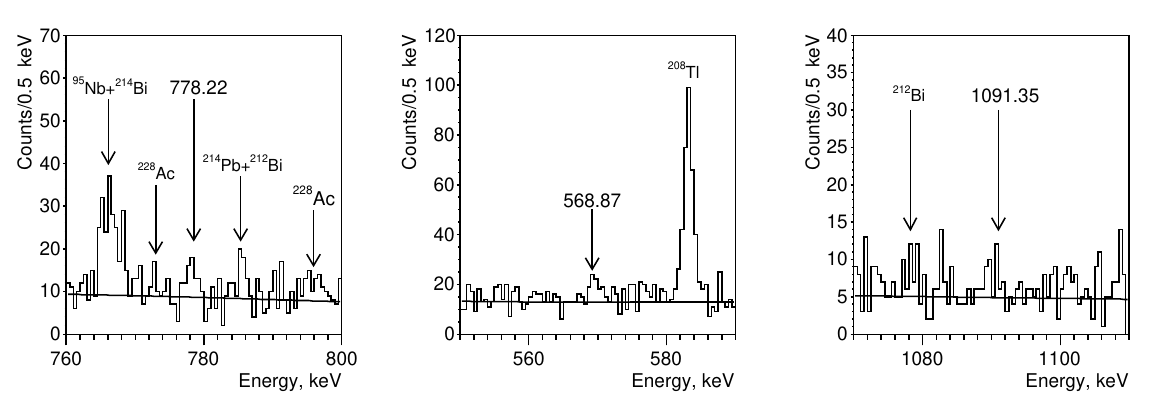}
\caption{Spectra of $\gamma$ quanta in the studied energy regions for 12625.34 h of measurements. The solid line is the background estimate (see the text of the paper).}
\label{fig:1}
\end{center}
\end{figure*}

The present work is devoted to the search for a single $\beta$ decay of $^{96}$Zr. The simplified decay scheme is shown in Fig. 1. The $^{96}$Zr (0$^+$) nucleus can undergo a single $\beta$ decay to $^{96}$Nb in the 4$^+$, 5$^+$ or 6$^+$ state. Theoretically, it was shown that the most probable transition is the one to $^{96}$Nb (5$^+$), the 4th forbidden unique transition. For example, within the Shell Model, the following estimates of half-lives were obtained for all three possible transitions \cite{ALA16}: $T_{1/2} (4^+) = 7.5\times10^{22}$ yr, $T_{1/2} (5^+) = 1.1\times10^{20}$ yr and $T_{1/2} (6^+) = 1.6\times10^{29}$ yr. After the decay of $^{96}$Zr to the 5$^+$ level of $^{96}$Nb, the $^{}$Nb nucleus transits to the ground state, emitting a $\gamma$ quantum with an energy of 44.19 keV (due to the low energy, this $\gamma$ quantum is not registered in our experiment). Then, this nucleus decays with a half-life of 23.35 hours to the excited states of $^{96}$Mo (96.7\% to the 5$^+$ excited state), emitting a cascade of $\gamma$ quanta (see Fig. 1). It is the registration of these $\gamma$ quanta that allows us to obtain information about the single $\beta$ decay of $^{96}$Zr. 

\begin{table*}
\caption{\label{tab:table1}Results of measurements for 12625.34 h for the three most intense $\gamma$ lines (778.22, 568.87 and 1091.35 keV).
 I$_{\gamma}$ is the intensity of the $\gamma$ line; 
$\epsilon_{\gamma}$ is the efficiency of registering one $\gamma$ quantum in a cascade; 
$\epsilon_{\gamma decay}$ is the efficiency of registering one $\gamma$ quantum per decay; N and B are the number of events and the background in the selected area; S is the signal in the selected area; S/$\surd B$ is the statistical significance of the signal; $\eta$ is the proportion of the Gaussian in the selected region; S$_{tot}$ is 
the total signal. }
\begin{ruledtabular}
\begin{tabular}{cccccccccc}
 Energy, keV  & I$_{\gamma}$, \%
& $\epsilon_{\gamma}$, \% & $\epsilon_{\gamma decay}$, \% & N & B & S (error)  & S/$\surd B$ & $\eta$ & S$_{tot}$ (error)\\ 
 & &    &  &  & &  &  & & \\ 
\hline
 778.22 & 96.45 & 1.183 & 1.141 & 94 & 60.52 & 33.48 (9.70) & 4.30 & 0.9158 & 36.56(10.59) \\
568.87  & 58.0 & 1.420 & 0.824 & 129 & 90.60 & 38.40 (11.36) & 4.03 & 0.9619 & 39.92(11.81) \\
1091.35  & 48.5 & 0.988 & 0.479 & 61 & 43.83 & 17.17 (7.81) & 2.59 & 0.9444 & 18.18(8.27) \\
 
\end{tabular}
\end{ruledtabular}
\end{table*} 

The experiment was carried out at the underground Baksan Neutrino Observatory of INR RAS (Baksan, Russia) at a depth of 4900 m w.e. using the low-background setup SNEG based on an HPGe detector. The detector volume is 211 cm$^3$, and the energy resolution is 2.0 keV (FWHM) at an energy of 1332 keV. The detector is surrounded by the layers of passive shielding: copper (180 mm), lead (150 mm), cadmium (1 mm) and borated polyethylene (50-80 mm). There is a cavity for placing samples near the detector. In order to remove the radioactive gas ($^{222}$Rn) present in the air from the working cavity, the internal volume of the shield is purged with the liquid nitrogen vapour from the Dewar vessel. The measurement was carried out with two samples of enriched Zr, specially produced for this experiment by the centrifugation method (atomic fraction of $^{96}$Zr is 88.28\%, total mass is 140.65 g). The enriched Zr also contains 11.42\% of $^{94}$Zr, 0.14\% of $^{92}$Zr, 0.04\% of $^{91}$Zr and 0.12\% of $^{90}$Zr. The first sample (located closer to the HPGe detector) is small pieces of zirconium boride with a total mass of 149.520 g; the mass of $^{96}$Zr is 90.38 g. The Zr content was provided by the manufacturer. Taking into account the zirconium content in the product, the most appropriate empirical formula for the compound is $^{96}$ZrB$_4$.
The second sample is ZrO$_2$ powder with a mass of 75.865 g; the mass of $^{96}$Zr is 50.27 g. Thus, the total mass of the studied $^{96}$Zr is 140.65 g (or $8.83\cdot 10^{23}$ nuclei of $^{96}$Zr). The bulk density of the first sample is 1.97 g/cm$^3$, and that of the second one is 1.00 g/cm$^3$. The enriched zirconium samples were packed in special containers made of ultra-pure material (nylon) using a 3D printer. The dimensions of the containers are $\oslash$ $85\times18.5$ mm. These containers were placed in the sample cavity close to the HPGe detector.

It should be noted that until recently the enriched $^{96}$Zr was obtained via electromagnetic separation. Therefore, the isotope was very expensive and only a few dozens of grams were produced worldwide. A few years ago, in Russia, the method was developed to obtain the enriched zirconium using centrifuges, and the first $\sim$ 180 g of $^{96}$Zr were produced for our experiment. In these measurements, only a part of the enriched zirconium was used. More information on the enriched zirconium samples can be found in \cite{ARE26}.

\begin{table*}
\caption{\label{tab:table1}Content of radioactive impurities in enriched Zr samples.}
\begin{ruledtabular}
\begin{tabular}{ccccccccc}
 Isotope & $^{238}$U ($^{234}$Th) & $^{238}$U ($^{234m}$Pa) & $^{226}$Ra & $^{235}$U & $^{232}$Th ($^{228}$Ac) & $^{40}$K & $^{137}$Cs & $^{60}$Co\\  \hline
 Activity, mBq/kg & $<$ 35.8 & $<$ 13.6 & $<$ 0.82 & $<$ 0.48 & $<$ 1.05 & $<$ 5.29 & $<$ 0.25 & $<$ 0.14 \\

\end{tabular}
\end{ruledtabular}
\end{table*} 

 Digitized signals are recorded for the subsequent pulse shape analysis. As a preprocessing step, a baseline correction for each waveform was performed by subtracting the mean value calculated from the initial portion of the signal, eliminating the ADC's constant offset. Then, a digital trapezoidal filter \cite{JOR94} was applied to the signal in the recursive form with specified parameters for the shaping time and plateau width, followed by double integration to obtain a trapezoidal response. To compensate for the exponential tail of the preamplifier, the pole-zero correction was used to ensure the plateau and to improve the stability of the amplitude estimate. The event amplitude was determined by averaging the signal values on the trapezoid plateau within the fixed window around its center.
\begin{equation}
    d(n)=v(n)-v(n-k)-v(n-l)+v(n-k-l),
    \label{eq:trapezoid}
\end{equation}

\noindent where $v(n)$ is the signal array;
$n$ is the counting number;
$l$ is the trapezoid shaping time;
$k$ is the "flat-top" width. 

The total energy spectrum in the energy region of 0-2800 keV recorded for 12625.34 h and the background spectrum (without samples) recorded for 2896.98 h are shown in Fig. 2. It can be seen that the background and the main spectrum are almost identical (taking into account the lower statistics of background measurements), which means that the concentration of radioactive impurities in the samples is quite low. All observed $\gamma$ rays have been identified as emitted from natural, cosmogenic and man-made radioactive isotopes. The most intensive lines are indicated in the figure. The $\gamma$ lines from $^{95}$Zr (724.20 keV and 756.73 keV) and $^{95}$Nb (765.80 keV) were detected in the main spectrum. This is due to the activation of Zr samples by neutrons while they were on the Earth's surface. $^{95}$Zr is formed as a result of the (n,2n) reaction with $^{96}$Zr and the (n,$\gamma$) reaction with $^{94}$Zr. $^{95}$Zr, with a half-life of 64 days, is converted via $\beta$ decay into $^{95}$Nb, which, in turn, is converted via $\beta$ decay ($T_{1/2}$ = 35 days) into the stable isotope $^{95}$Mo. As a result, these peaks (from $^{95}$Zr and $^{95}$Nb) were formed during the first months of measurements, and then they stopped growing. Figure 3 shows the $\gamma$-quanta spectra after 12625.52 h of measurement in the energy range of interest. The figure shows the desired peaks from the $\beta$ decay of $^{96}$Zr and the identified background peaks associated with natural radioactivity of surrounding materials and samples. The peaks at energies of 778.22, 568.87 and 1091.35 keV, characteristic of the beta decay of $^{96}$Zr, are clearly visible. The solid line is the background estimate calculated using the ROOT software package. The three most intense $\gamma$ lines were studied: 778.22 keV (96.45\% per decay), 568.87 keV (58.0\%) and 1091.35 keV (48.5\%). Using adjacent background $\gamma$ lines (583 and 911 keV), an estimate of the average energy resolution of each studied $\gamma$ line was made (over the entire measurement period). The resolution (FWHM) was established as $(1.70 \pm 0.14)$, $(1.97 \pm 0.17)$ and $(2.13 \pm 0.19)$ keV for 568.87, 778.22 and 1091.35 keV, respectively.
 The selected areas for each peak were identified as (567.5-571.0), (776.5-780.0) and (1089-1093.5) keV covering almost all useful events ($\sim$ 92-96\%). Table 1 shows the measurement results for these three lines with the registration efficiency of the corresponding $\gamma$ quanta. The efficiencies were calculated using the Monte Carlo method. The calculations were performed by two independent groups using the Geant3.21 and Geant4 software. Both groups have obtained almost identical results. The HPGe detector was calibrated using a source of known activity (a bulk $^{238}$U source with an activity of $1555 \pm 8$ Bq \cite{BAR26}). The Monte Carlo calculations agree with the calibration results within (3-5)\%. The sum of the three peaks gives S$_{tot}$ = (94.66 $\pm$ 17.89) events. Considering the total registration efficiency of 2.444\% and the measurement time of 12625.34 h, the half-life of $^{96}$Zr $\to$ $^{96}$Nb can be estimated as:

\begin{equation} 
 T_{1/2} = 2.27^{+0.53}_{-0.36}\times 10^{20} yr.
\end{equation}
 
Only the statistical error is given. Considering the excess of signal over the background for all three lines (see Table 1), the statistical significance of the obtained result can be estimated as 6.4 $\sigma$. If we take into account the systematic error, the final result is as follows:
 
 \begin{equation}
   T_{1/2} = [2.27^{+0.53}_{-0.36}(stat) \pm 0.27(syst)]\times10^{20} yr.
 \end{equation} 
  
The main contribution to the systematic error is made by uncertainties in the calculated values of the registration efficiency ($\pm 5\%$), in the determination of the average background ($\pm 10\%$) and in the assessment of the total number of events S$_{tot}$ ($\pm 3\%$; associated with uncertainties in energy resolution). A small contribution to the effect can also be made by the hypothetical decay of $^{96}$Zr into the 0$^+_1$ excited state of $^{96}$Mo, which is accompanied by a cascade of $\gamma$ quanta with energies of 369.80 and 778.22 keV. According to theoretical estimates, the half-life of this transition is $(2.4-3.8)\times10^{21}$ years \cite{STO95,STO96,TOI97}. Thus, the contribution of this decay to the 778.22 keV line should be $\sim$ 2-3 events. Ultimately, this leads to an additional systematic error (+3\%) in the half-life value.

The result can be compared with the best existing experimental limit $T_{1/2} > 3.8\times 10^{19}$ years \cite{ARP94}. Theoretical calculations within the QRPA model predict a value of $2.4\times10^{20}$ yr \cite{HEI07}, while within the Shell Model they predict $1.1\times10^{20}$ yr \cite{ALA16} and $1\times10^{20}$ yr \cite{KOS20}. It can be seen that the value we have obtained is in good agreement with the QRPA prediction and approximately 2 times higher than the one in the predictions of the Shell Model. 
 
   The resulting spectrum was also analyzed for the content of radioactive impurities in the sample (also using the background spectrum collected over 2896.98 hours). The results are presented in Table 2. Only upper limits were obtained. One can conclude that the samples of enriched Zr are quite pure.

Thus, for the first time, the single $\beta$ decay of $^{96}$Zr with a half-life of $T_{1/2} = [2.27^{+0.53}_{-0.36}(stat) \pm 0.27(syst)]\times10^{20}$ yr has been registered.
In our experiment, we cannot distinguish between beta transitions to 4$^+$, 5$^+$ or 6$^+$ levels of $^{96}$Nb. But most likely, this is beta decay to the 5$^+$ excited state of $^{96}$Nb.
It must be emphasized that $^{96}$Zr has the longest half-life among all known $\beta$ decay nuclei and the decay recorded is one of the rarest beta decays recorded so far. More rare beta decay was observed in $^{115}$In. Although the half-life of $^{115}$In is $\sim 4.4\times10^{14}$ years (transition to the ground state of $^{115}$Sn), decay to the first excited state of the daughter nucleus (497.33 keV) with a half-life of $\sim 4.3\times10^{20}$ years has also been recorded (see, for example, \cite{AND11}).
Our measurements can also be used to search for the $2\nu\beta\beta$ decay of $^{96}$Zr to the first 0$^+$ excited state of $^{96}$Mo. However, this will be a subject of a separate publication. If this type of decay is registered and the corresponding half-life is measured, an even more unique context will arise for the theoretical study of this nucleus. The calculations must be consistent considering three decays: the 4th forbidden ordinary $\beta$ decay and two transitions via $2\nu\beta\beta$ decay (to the ground and excited states of the daughter nucleus). This allows us to hope for a more precise determination of parameters of the theory, which will improve the accuracy of NME calculations for both $2\nu\beta\beta$ decay and $0\nu\beta\beta$ decay. Moreover, the study of the $^{96}$Zr system will help to clarify the situation with "quenching" of the axial vector constant $g_A$.

\begin{acknowledgments}
Supported within the State Project "Science" by the Ministry of Science 
and Higher Education of the Russian Federation (075-15-2024-541).
\end{acknowledgments}


\end{document}